\documentclass[nolinenumbers]{aastex631}

\begin{document}

\title{A Study on the X-ray pulse profile and spectrum of the Crab pulsar \\
using {\sl NICER} and {\sl Insight-HXMT}'s Observations}

\correspondingauthor{Lin-Li Yan, You-Li Tuo}
\email{yan.linli@foxmail.com, tuoyl@ihep.ac.cn}

\author[0000-0002-2244-4222]{Lin-Li Yan}
\affiliation{School of Mathematics and Physics, Anhui Jianzhu University,
Hefei, Anhui 230601, China}
\affiliation{Key Laboratory of Architectural Acoustics Environment
of Anhui High Education Institutes,
Hefei, Anhui 230601, China}
\affiliation{Key Laboratory of Advanced Electronic Materials and Devices,
Anhui Jianzhu University,
Hefei, Anhui 230601, China}

\author[0000-0003-3127-0110]{You-Li Tuo}
\affiliation{Key Laboratory of Particle Astrophysics,
Institute of High Energy Physics, Chinese Academy of Sciences,
Beijing 100049, China}

\author[0000-0002-2749-6638]{Ming-Yu Ge}
\affiliation{Key Laboratory of Particle Astrophysics,
Institute of High Energy Physics, Chinese Academy of Sciences,
Beijing 100049, China}

\author[0000-0003-3248-6087]{Fang-Jun Lu}
\affiliation{Key Laboratory of Particle Astrophysics,
Institute of High Energy Physics, Chinese Academy of Sciences,
Beijing 100049,  China}
\affiliation{Key Laboratory of Stellar and Interstellar Physics,
School of Physics and Optoelectronics, Xiangtan University,
Xiangtan, Hunan 411105, China}

\author{Shi-Jie Zheng}
\affiliation{Key Laboratory of Particle Astrophysics,
Institute of High Energy Physics, Chinese Academy of Sciences,
Beijing 100049, China}

\author{Ling-Jun Wang}
\affiliation{Key Laboratory of Particle Astrophysics,
Institute of High Energy Physics, Chinese Academy of Sciences,
Beijing 100049, China}

\begin{abstract}
We analyze the energy dependence of the X-ray pulse
profile and the phase-resolved spectra (PRS) of the Crab
pulsar using observations from the Neutron star Interior Composition
Explorer ({\sl NICER}) and the Hard X-ray Modulation Telescope
({\sl Insight-HXMT}). We parameterize the pulse profiles and quantify
the evolution of these parameters in a broad energy band of 0.4-250 keV.
A log-parabola function is used to fit the PRS in 2-250 keV,
and the curvature of the spectrum, i.e., the evolution of photon
index with energy, as represented by the parameter $\beta$ of
the log-parabola model also changes with phase.
The relation of $\beta$ and phase have two turning points
slightly later than those of the pulse intensity profile, where
the values of the $\beta$ are the lowest, suggesting that the
energy loss rate of the particles are the lowest in the corresponding regions.
A three-segment broken power-law model is also used to fit those PRS.
The differences between the hard spectral index and the soft ones
have distribution similar to that of $\beta$, confirming the fitting
results of the log-parabola model, while the broken energies are
generally higher in the region bridging the two pulses.
We find anti-correlations between the spectral indices and curvature
of the log-parabola model fitting and similar anti-correlation between
the spectral indices and broken energies of the broken power law model
fitting, suggesting a scenario that the highest energy particles
are produced in regions where the radiation energy loss
is also the strongest.
\end{abstract}


\section{Introduction}
The Crab pulsar (PSR B0531+21) is one of the most frequently studied
objects in X-ray astrophysics. It has a period of about 33 ms, a spin
down rate of $\dot{P}=4.2\times10^{-13}$ s s$^{-1}$ and an energy loss
power of $\dot{E}=5\times10^{38}$ erg s$^{-1}$.
Although observational data on this source have been gathered for
about 50 years, the physics involved in this pulsar has not yet been
completely understood. The position and size of the radiation regions,
the distribution of secondary pairs, the magnetic field, the inclination
and viewing angles, etc., together determine the pulse profiles
and spectra in different energy bands \citep{Cheng1986a, Cheng1986b}.
Studying the pulse profile and spectrum variations will thus
help to explore the links between the multi-wavelength radiation
properties and the physical conditions on this pulsar.

The pulse profiles of the Crab pulsar at all wavelengths are
dominated by two pulses, separated by about 0.4 pulse phase. Previous
studies show that the exact pulse morphology varies as a function of
photon energy \citep{Eikenberry1997ApJ, Kuiper2001AA, Ge2012ApJS, Tuo2019RAA}:
the intensities of both the second pulse and the bridge emission
increase with energy relative to that of the first pulse in X-ray and
followed by a downward trend above 1 MeV \citep{Kuiper2001AA}.
However, some of these studies were either done in narrow energy
bands that can not give quantitative results \citep{Eikenberry1997ApJ, Tuo2019RAA},
or incomplete for only adopting the flux ratio of the two pulses to
describe the evolution of the pulse profile \citep{Kuiper2001AA}.

Actually the information contained in the energy dependence of the
pulse profile can be revealed in more detail by phase resolved X-ray
spectroscopy. The overall X-ray spectrum of the Crab pulsar could be
well described by a power-law function and its spectral index varies
with energy \citep{Massaro2000AA, Weisskopf2004ApJ, Mineo2006AA}.
The phase-averaged total spectrum of the Crab nebula and pulsar
in 1-100 keV has a photon index $\Gamma\sim2.1$ (RXTE, BeppoSAX, EXOSAT,
INTEGRAL/JEM X, \citet{Kirsch2005}), and a softer index of $\sim2.20-2.25$
above 100 keV (INTEGRAL/SPI/ISGRI, CGRO \citet{Mineo2006AA, Kuiper2001AA}).
The pulsed spectrum has a more complex evolution \citep{Kuiper2001AA} and the
photon index is about 1.6 in 0.3-3.8 keV \citep{Weisskopf2011ApJ},
1.81 in 5-60 keV, 1.91 in 15-250 keV, and 1.96 in 0.1-300 GeV \citep{Yan2018}.
For the phase-resolved X-ray spectra, \citet{Pravdo1997} found a phase
evolution of the spectral index across the X-ray pulse in a
reverse $S$ shape in 5-200 keV, i.e., the spectral index increases until the
peak of the first pulse and then falls off, while the
spectral index keeps increasing from the leading wing to the trail wing of the
second pulse. Such a phase evolution trend holds albeit with different values
in 0.3-3.8 keV \citep{Weisskopf2011ApJ}, 1-10 keV \citep{Vivekanand2021AA},
3-60 keV \citep{Ge2012ApJS}, 3-78 keV \citep{Madsen2015APJ},
11-250 keV \citep{Tuo2019RAA}, 15-500keV \citep{Li2019},
20-500 keV \citep{Mineo2006AA} and $>$ 100MeV \citep{Abdo2010}.

A three-dimensional outer magnetospheric gap model was proposed to
explain the pulse profile, phase-resolved spectra, and polarization of
the Crab pulsar \citep{Cheng2000, Takata2007ApJ, Tang2008ApJ}.
Specifically, in the model of \citet{Cheng2000}, the X-ray photons
at different phases are produced by the radiations from different height
range of regions in the magnetosphere.
Recently, particle-in-cell (PIC) simulations are developed to
study the pulsar magnetosphere that is more complex than a simple
dipole \citep{Philippov2014, Philippov2015ApJa, Philippov2015ApJb}.
Such PIC methods can therefore produce more complicated pulse profiles
and have been successfully applied to the interpretation of Fermi
observations of pulsars \citep{Cerutti2016MNRAS, Philippov2018ApJ}.
As the X-ray spectrum reflects the energy distribution and radiation
cooling of the high energy electrons, phase-resolved spectroscopy
in a broad X-ray band can provide information for theoretical
studies on the high energy particle production and cooling, as well
as the magnetic field geometry of the pulsar magnetosphere.

In this work, we use observations from the Neutron star Interior
Composition Explorer ({\sl NICER}) and the Hard X-ray Modulation
Telescope ({\sl Insight-HXMT}) to analyse the pulse morphology
and phase-resolved spectra (PRS) for the Crab pulsar, providing
further constraints on models of pulsar X-ray emission.
The organization of this paper is as follows: data processing and
reduction are presented in Section 2, analysis methods in
Section 3, results in Section 4, discussions on the physical
implications of our results and a summary in Section 5 and
Section 6, respectively. Throughout the paper, errors of the parameters
are at the one standard deviation level.

\section{Observations and Data Reductions}
\label{sect:Obs}
\subsection{Timing Ephemeris from Jodrell Bank Observatory}
A 13 m radio telescope at Jodrell Bank Observatory monitors the Crab
pulsar daily \citep{Lyne1993MNRAS}, offering a radio ephemeris
(denoted as JBE) that is used for the analyses of {\sl NICER}
and {\sl Insight-HXMT} data.
For monthly validity intervals, this ephemeris contains up-to-date
parameters of the rotation frequency and its first two time
derivatives. The database is available through the World Wide
Web (http://www.jb.man.ac.uk/pulsar/crab.html).
These JBE parameters, the pulsar coordinate R.A.=$83^{\circ}.63322$,
decl.=$22^{\circ}.01446$(J2000) \citep{Abdo2010}, and
solar system ephemeris JPL DE200 are used in this work.

\subsection{{\sl NICER} Observations and Data Reductions}
The {\sl NICER} is an International Space Station payload devoted to
the study of neutron stars through soft X-ray timing in 0.2-12 keV.
Its X-ray Timing Instrument (XTI) is an aligned collection of 56
X-ray concentrator optics (XRC) and silicon drift detector (SDD) pairs.
Each XRC collects X-rays over a large geometric area from a roughly
30 arcmin$^2$ region of the sky and focuses them onto a small SDD.
The SDD detects individual photons, recording their energies with
good spectral resolution and their detection times to a $\sim$100
nanoseconds RMS relative to Universal Time \citep{Prigozhin2016SPIE}.

The {\sl NICER}'s observations used in this paper are from 2017 August
25 to 2019 November 26,  but only those with available JBE are used,
and the total exposure time is about 339 ks.
The data reduction is carried out by using NICERDAS software version 7a
included in HEAsoft version 6.27.2, with calibration database version 20200722.
Standard calibration processes as well as filtering to an entire NICER
observation are  performed by the command \texttt{nicerl2}. The cleaned photon
events produced by the previous step are corrected to the solar system
barycenter (SSB) and then used to fold pulse profiles in different soft
X-ray energy bands. The cleaned events after corrected to SSB are also used
to get the phase resolved spectrum. We extract the source spectrum with
HEASoft XSELECT package (V.2.4e).

\subsection{{\sl Insight-HXMT} Observations and Data Reductions}
The {\sl Insight-HXMT} is China's first X-ray astronomy satellite.
It has three main payloads: the High Energy X-ray telescope (HE,
$20-350\,\rm{keV}$, $5100\,\rm{cm}^2$), the Medium Energy
X-ray telescope (ME, $8-35\,\rm{keV}$, $952\,\rm{cm}^2$) and the
Low Energy X-ray telescope (LE, $1-12\,\rm{keV}$, $384\,\rm{cm}^2$) \citep{Zhang2020HXMT}.
Only the HE and ME observations of the Crab pulsar with available
JBE are used in this work, because of their high time resolution
($\sim$ 2\,$\mu$s for HE and $\sim$ 20\,$\mu$s for ME)
and their complementarity to the {\sl NICER} data.
All data from these two instruments are reduced via data processing
pipeline \texttt{hpipeline} included in HXMTsoft (v2.04).
Events are corrected to SSB by using the command \texttt{hxbary}.
All the data is collected in MJD 57992-59228 (2017 August 27 to 2021
January 14) and in the good time intervals (GTIs) recommended by pipeline.
The selected observations have a total exposure of about 596 ks for HE
and 758 ks for ME.

\section{Data Analysis}
\label{sect:data}
\subsection{Normalization of Pulse Profiles}
\begin{figure}[ht!]
\begin{center}
\includegraphics[angle=0, width=0.6\textwidth]{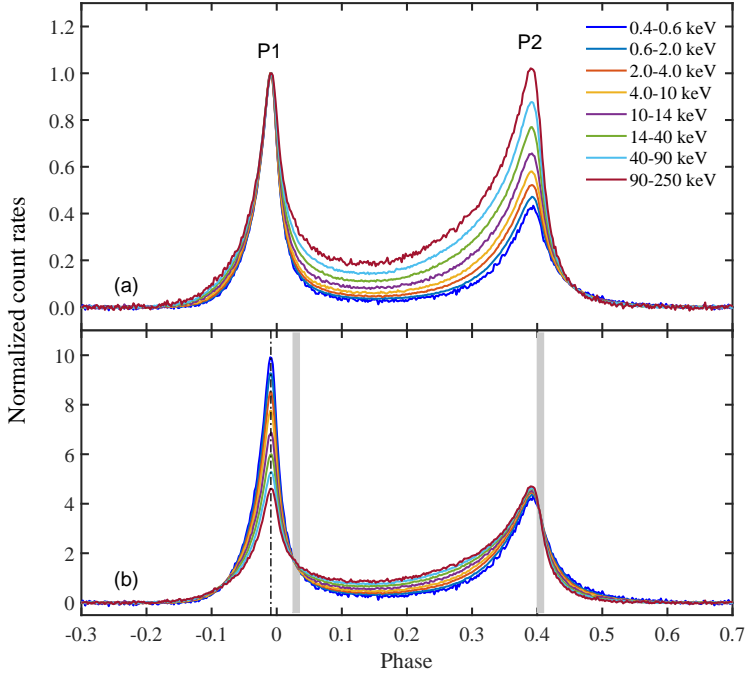}
\caption{The normalized profiles of the Crab pulsar in different
energy bands from {\sl NICER} and {\sl Insight-HXMT}'s observations.
The whole phase range is divided into 500 bins. The profiles
in 0.4--10\,keV are from {\sl NICER}'s observations, and in 10--250\,keV
from {\sl Insight-HXMT}'s observations. In panel (a), profiles
are normalized by the peak intensity of P1 with the off-pulse
emission subtracted; In panel (b), profiles are normalized by
the mean count rates of all phase bins with the off-pulse emission
subtracted. The dash-dotted line marks the location of the P1,
and two gray bars (phase 0.025--0.035, and 0.40--0.41) mark the
turnover phase positions of the spectral index variation trends.
\label{diff_energy_profiles}}
\end{center}
\end{figure}

Figure \ref{diff_energy_profiles} shows the pulse profiles in
different energy ranges, in which we denote the first pulse that
is more prominent in soft X-rays as P1, and the second pulse as P2.
To show the difference between the profiles more clearly,
all the original profiles are first subtracted by their respective
mean count rate in the off-pulse region (phase 0.6-0.8; \cite{Yan2018})
and then normalized by the peak flux of P1. After this process, the
peak values of P1 for all the profiles equal to 1, as shown in Figure
\ref{diff_energy_profiles}(a).
The other normalization method is also used in this work: every
profile in an energy band is subtracted by its off-pulse count
rate and then divided by the mean photon count rate in the whole
period. The profiles normalized by the second method make it
easily to observe the overall energy evolution trend of the
pulse profiles. The location of the maximum of P1 is aligned as phase
about -0.01 by using JBE.

For observations from {\sl NICER}, because of the high detector
noise in lower energy, only profiles in 0.4-12 keV are  considered,
according to the suggestion at
https://heasarc.gsfc.nasa.gov/docs/nicer/data\_analysis/nicer\_analysis\_tips.html.
For the observations from {\sl Insight-HXMT},  the HE profiles are
in 30-250 keV and the ME profiles are in 8-30 keV.

\subsection{Parameterization of the Pulse Profiles}
\label{sect:parameters}
In order to study the energy dependence of the pulses more
comprehensively, we describe the shape of a pulse profile with
five parameters, which are the ratio of the peak intensity
of P2 to that of P1 ($R_{I}$), the separation of the two
pulses ($\Delta{\Phi}$), the full width at half maximum (FWHM)
of P1 ($W_{1}$), the FWHM of P2 ($W_{2}$), and the ratio of the
fluxes within the FWHM phase ranges ($R_{F}$) of the two pulses.
Because the shapes including the peak phases of the two pulses
change with energy, the evolution of $R_{F}$ with energy differs
slightly from that of $R_{I}$. The peak intensities and phase
positions of the two pulses are obtained by using the empirical
formula proposed by \citet{Nelson1970}, as introduced in \citet{Ge2012ApJS}.

To explore the exact profile evolution of the pulses from another
aspect, we calculate the flux ratios between the left side and the
right side of P1 and P2 in different energy bins. The four phase
intervals from which the fluxes are calculated are: LP1 (from the
left FWHM phase to the peak of P1), RP1 (from the peak of P1 to
its right FWHM phase to the peak of P1), LP2 (from the left
FWHM phase to the peak of P2), RP2 (from the peak of P2 to its right
FWHM phase to the peak of P2).

\subsection{Spectral Fitting}
The XSPEC (v.12.10.0c) in the software package HEASOFT is used
to fit the spectra created in Section \ref{sect:Obs}. The
PRS of the Crab pulsar extracted from the {\sl NICER} observations
were fitted by a power law model by \citet{Vivekanand2021AA},
and the similar spectral analyses of the {\sl Insight-HXMT} data
were also carried out by \citet{Tuo2019RAA}.
In this work, we fit the PRS from {\sl NICER} and {\sl Insight-HXMT}
data jointly, with the log-parabola model \citep{Massaro2000AA} and
the broken power law models, respectively.

\begin{figure}[ht!]
\centering
\includegraphics[width=0.5\textwidth,angle=-90]{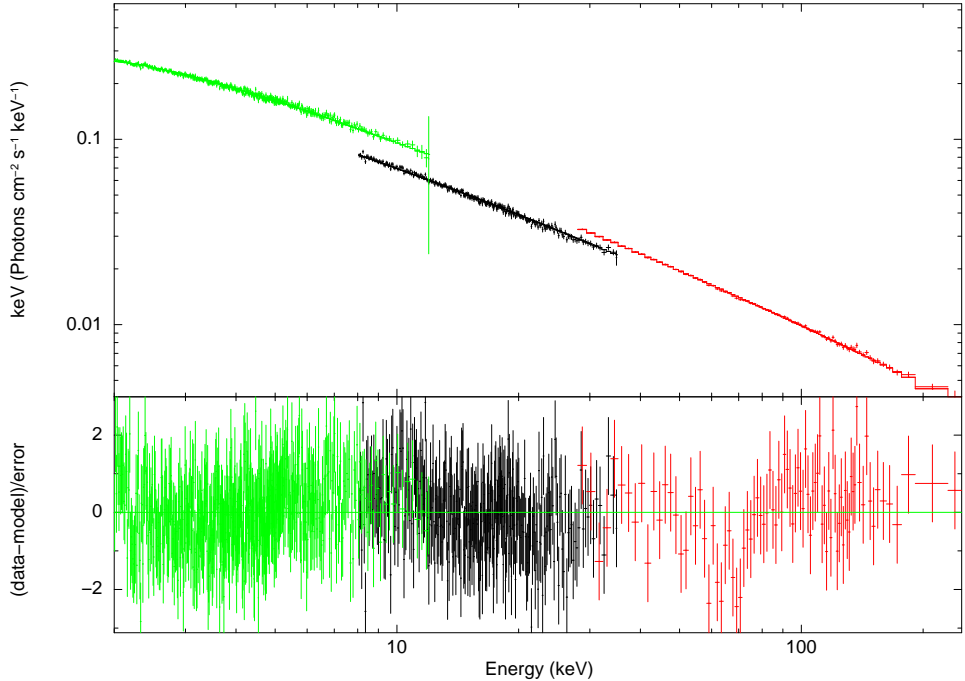}
\caption{The joint fitting of phase average spectra for {\sl{NICER}} and
{\sl{Insight-HXMT}}'s observations by using the \textit{logpar} model.
The green points are the {\sl{NICER}} data in 2--12\,keV, the black ones are
the HXMT-ME data in 8--35\,keV, the red points are the HXMT-HE data
in 27--250\,keV. The upper and lower panels show the spectra and residuals
in terms of sigmas with error bars, respectively.}
\label{fig:spec_average}
\end{figure}

The log-parabola model involves three components, \textit{wabs*constant*logpar},
among which \textit{wabs} accounts for the interstellar medium absorption
whose column density $N_H$ is fixed as 0.36$\times10^{22}$ cm$^{-2}$ \citep{Ge2012ApJS},
\textit{constant} is used to compensate the mis-match of the effective areas
of the three instruments (i.e. XTI, ME, and HE), and \textit{logpar}
is a power-law with an index which varies with energy $E$ (eq. \ref{eq:1})
to show the detailed energy evolution of the spectrum $F(E)$ \citep{Massaro2000AA}
in a broad band,
\begin{equation}
F(E)=K(E/E_{0})^{-(\alpha+\beta{log(E/E_{0})})}.
\label{eq:1}
\end{equation}
where $E_{0}=$1 keV, and \textit{$\alpha$}, \textit{$\beta$} and the
normalization value $K$ are set as free parameters to fit. \textit{$\alpha$}
corresponds to the photon index at 1 keV, and \textit{$\beta$} is used to
represent the curvature of the spectrum.
The \textit{constant} for the {\sl NICER} spectra is fixed at 1,
while for the ME and HE spectra, their \textit{constant} are determined
by jointly fitting the phase averaged spectra of the three instruments.
As shown in Figure \ref{fig:spec_average}, the phase averaged spectra from
these three instruments could be well fitted by the \textit{logpar}
model, and the obtained parameters are: \textit{constant} equal to
$0.623\pm0.002$ and $0.831\pm0.004$ for ME and HE, respectively,
$\alpha=1.581\pm0.004$, $\beta=0.109\pm0.001$, $K=0.471\pm0.002$,
and the reduced $\chi^2=0.742$ for degree of freedom of 1607.
Specifically, the two \textit{constant} values are used when we fit
the PRS.

\begin{figure}[ht!]
\centering
\includegraphics[width=0.7\textwidth]{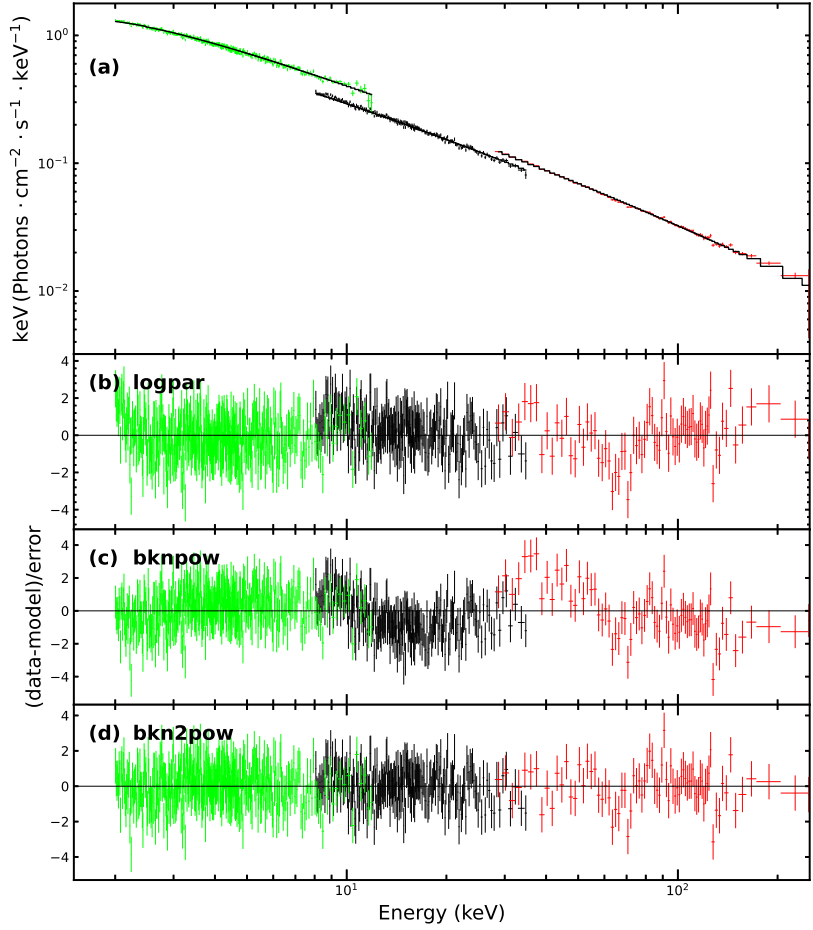}
\caption{The joint fitting for {\sl{NICER}} and {\sl{Insight-HXMT}}'s
observations in phase 0--0.01. Panel (a) shows the broad band unfolded
spectra, the fitting residuals of three models, \textit{logpar},
\textit{bknpow}, and \textit{bkn2pow}, are shown in panel (b), (c) and (d),
respectively. The green points are the {\sl{NICER}} data in 2--12\,keV,
the black ones are the {\sl{HXMT}}-ME data in 8--35\,keV, the red points
are the {\sl{HXMT}}-HE data in 27--250\,keV.}
\label{fig:spec_3models}
\end{figure}

We also used a two-segment broken power law function (\textit{bknpow})
and a three-segment broken function (\textit{bkn2pow}) to search for
the turnover energy of a spectrum of the Crab pulsar. Again, a \textit{wabs}
model is used to account for the absorption of the interstellar medium
and \textit{constant} to adjust the mis-match between the instruments.
Figure \ref{fig:spec_3models} shows the fitting results
of a spectrum in phase 0.00--0.01 by using the \textit{wabs*constant*bknpow}
and \textit{wabs*constant*bkn2pow} models.
Compared with the \textit{bknpow} model, \textit{bkn2pow} gives a
much better fit to the data and so the fitting results of $bkn2pow$
model are discussed later in this work.

One spectrum is created for each phase bin with a size of 0.01 in
phase -0.12 to 0.5. These spectra are fitted with the \textit{logpar} model
each. However, some of the spectra are merged into one when fitted with
the \textit{bkn2pow} model in order to improve the statistics.
The background spectrum is created using all the events in phase range
0.6-0.8.

\section{Results}
\label{sect:results}
\subsection{Evolution of the Pulse Profiles with Energy}
Our analyses present the evolution of the pulse profiles with
energy much more quantitatively than in the previous studies.
As shown in Figure \ref{diff_energy_profiles}, the peak
intensity ratio and the FWHM of the two pulses all manifest
an increasing trend with energy.
The intensity of P1 decreases with energy and this trend
reverses at phase about 0.035. The intensity of the ``bridge"
region between this phase and 0.4 increases with energy,
and then the intensity starts to decrease with energy again,
as displayed in Figure \ref{diff_energy_profiles}(b).

\begin{figure}[ht!]
\begin{center}
\includegraphics[angle=0, width=0.6\textwidth]{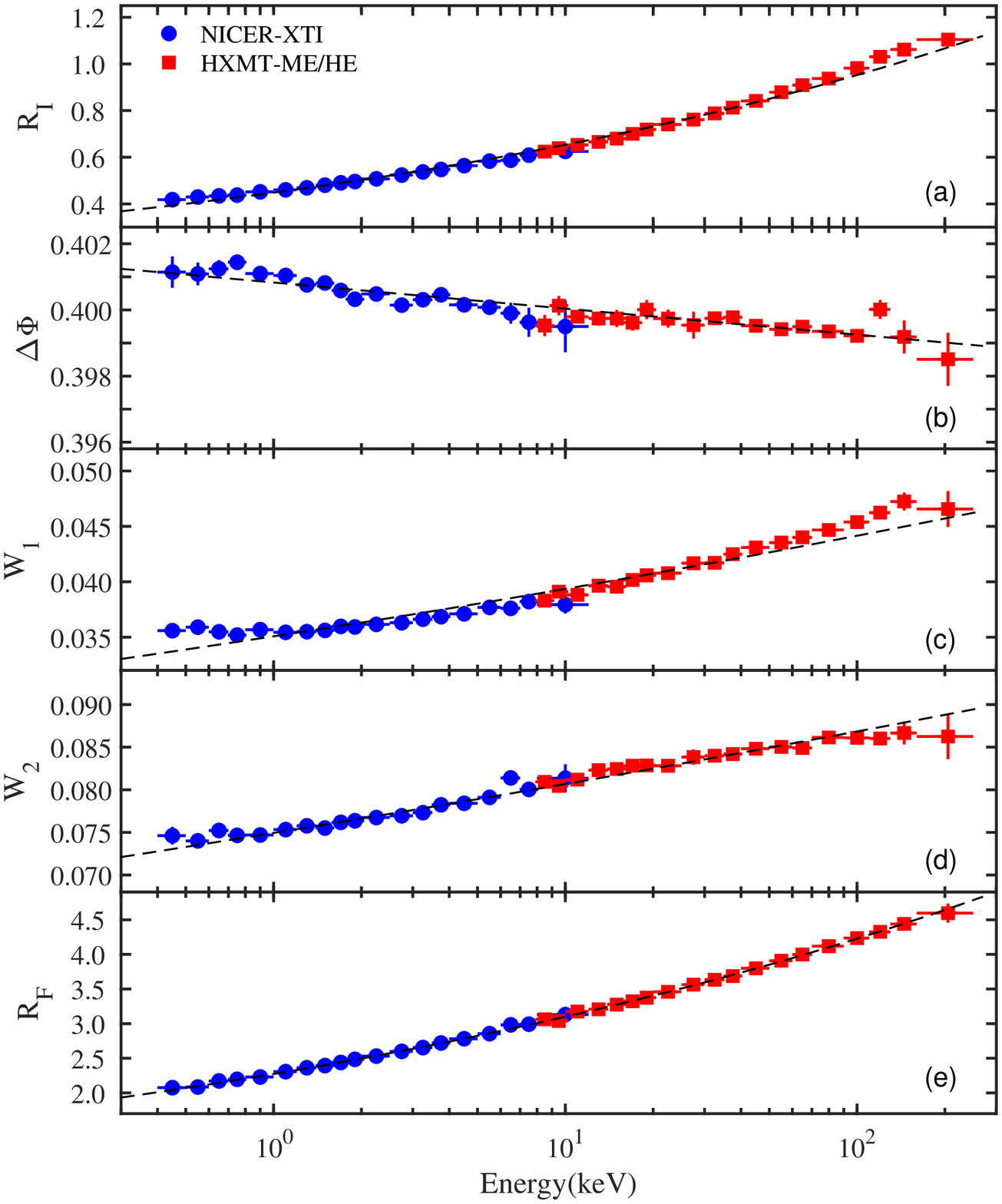}  
\caption{The pulse shape parameters of the Crab pulsar in X-ray.
$R_{I}$ is the intensity ratio of two pulses. $\Delta{\Phi}$ is the
phase separation of two pulses. $W_{1}$ and $W_{2}$ represent the FWHM
of P1 and P2, respectively. $R_{F}$ is used to denote the flux ratio
of the two pulses within FWHM. All the shape parameters evolves
with energy gradually, and the black dashed lines represent the joint
fitting results of data points from the two satellites with a power
law function respectively.
\label{shape_parameters}}
\end{center}
\end{figure}

Figure \ref{shape_parameters}(a) shows that the intensity
ratio of two pulses. $R_{I}$ increases monotonously and becomes
higher than 1 above 120 keV.
The separation between the two pulses decreases with energy in general,
but might increase with energy below 0.8 keV (Figure \ref{shape_parameters}(b)).
The pulse widths shown in panels (c) and (d) increase with energy,
albeit with different slopes.
The flux ratio of two pulses ($R_{F}$) also increases with energy,
from $2.08\pm0.03$ at 0.45 keV to $4.60\pm0.14$ at about 200 keV,
with a slightly larger gradient than that of $R_{I}$ (panel(e)).

In order to quantify the evolutionary trends of the above parameters,
we try to fit the data points with different models and find that
the power law function $Y=c_{1}*E(keV)^{c_{2}}$ can give an acceptable fit,
where $Y$ represents different shape parameters,
and $c_{1}$ and $c_{2}$ are coefficients of the evolution. All the data
points from {\sl{NICER}} and {\sl{Insight-HXMT}} are fitted, and the
parameters are listed in Table \ref{table:1}.

\begin{table}[ht!]
\caption{The fitted results of the pulse shapes with a power law model.}
\footnotesize
\label{table:1}
\medskip
\begin{center}
\begin{tabular}{c c c c }
\hline \hline
&$Y$           &$c_{1}$          &  $c_{2}$          \\
&                & $\times{10^{-2}}$ & $\times{10^{-2}}$   \\
\hline
& $R_{I}$        & $44.84\pm0.18$    & $16.36\pm0.18$    \\
& $\Delta{\Phi}$ & $40.083\pm0.005$  & $-0.086\pm0.005$  \\
& $W_{1}$        & $3.51\pm0.01$     & $5.00\pm0.19$     \\
& $W_{2}$        & $7.49\pm0.01$     & $3.20\pm0.06$     \\
& $R_{F}$        & $227.35\pm0.23$   & $13.47\pm0.05$    \\
\hline
& $R_{F,P2/P1}$      & $93.47\pm0.11$ &$13.78\pm0.05$   \\
& $R_{F, Bridge/P1}$ & $23.60\pm0.09$ &$27.92\pm0.12$   \\
\hline \hline
\end{tabular}
\end{center}
\end{table}

\begin{figure}[ht!]
\begin{center}
\includegraphics[angle=0, width=0.6\textwidth]{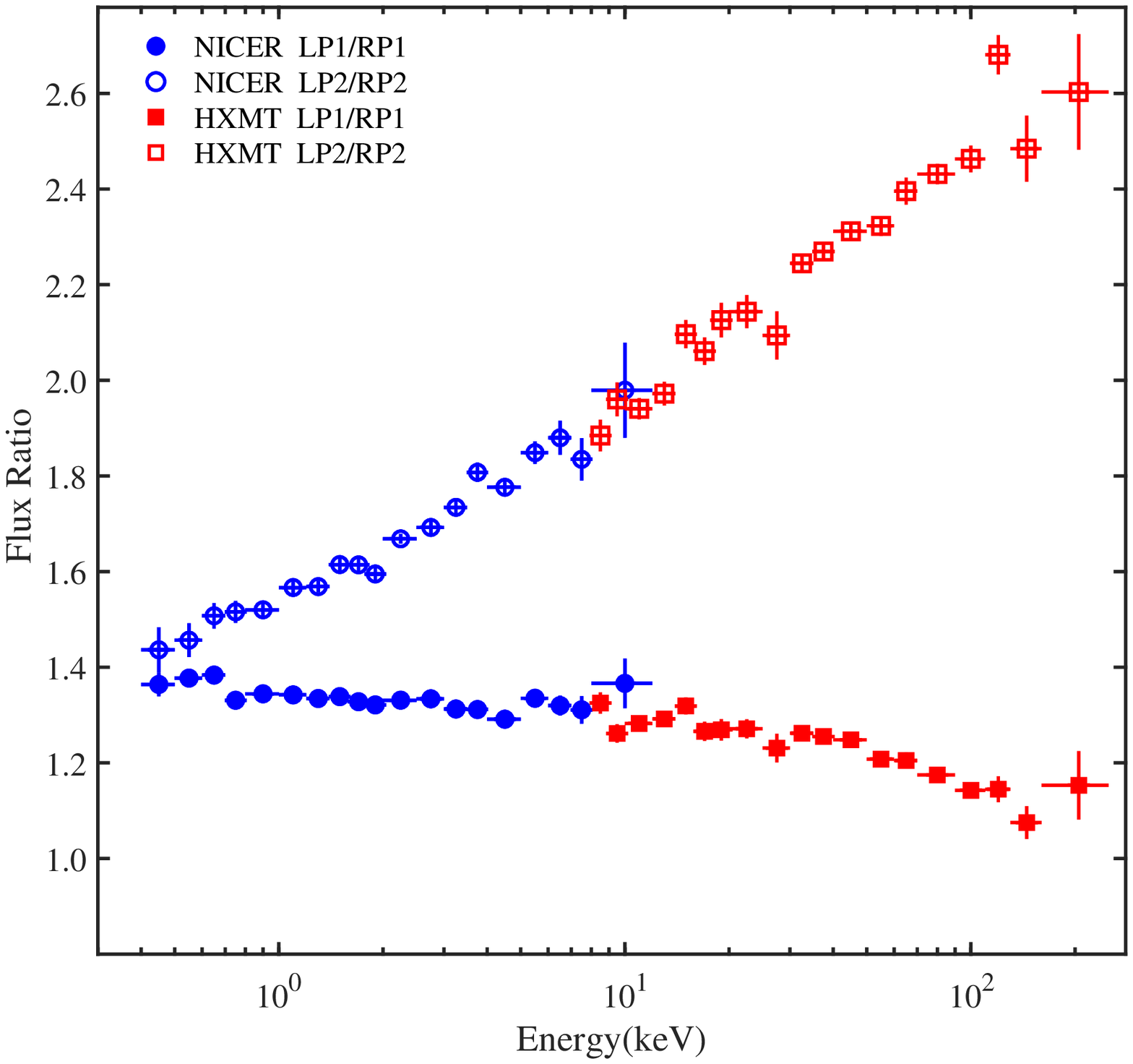}
\caption{ The flux ratio between the left side and the right
side of P1 and P2. The phase definitions are declared in Section 3.2.
\label{shape_parameters02}}
\end{center}
\end{figure}

Flux ratios of the left side to the right side are also obtained
for the two pulses respectively to  represent the evolution of
their shapes with energy, which are given in Figure \ref{shape_parameters02}.
Combining with the results in Figure \ref{shape_parameters},
we find that as the photon energy increases,
(1) the flux of P1 relative to the entire flux drops, leading to
a broader pulse and a lower height peak; the flux ratio between
the left side and the right side of P1 decreases slightly;
(2) the flux ratio of P2 relative to the entire flux rises, leading
to a broader pulse and a higher peak height; the flux ratio between
the left side and the right side of P2 increases obviously;
(3) the flux of the bridge region relative to the entire flux rises;
(4) the change trend of pulse intensity turns over at the
phases about 0.035 and 0.4, which are not aligned with the peak
positions of P1 and P2.

\begin{figure}[ht!]
\begin{center}
\includegraphics[angle=0, width=0.6\textwidth]{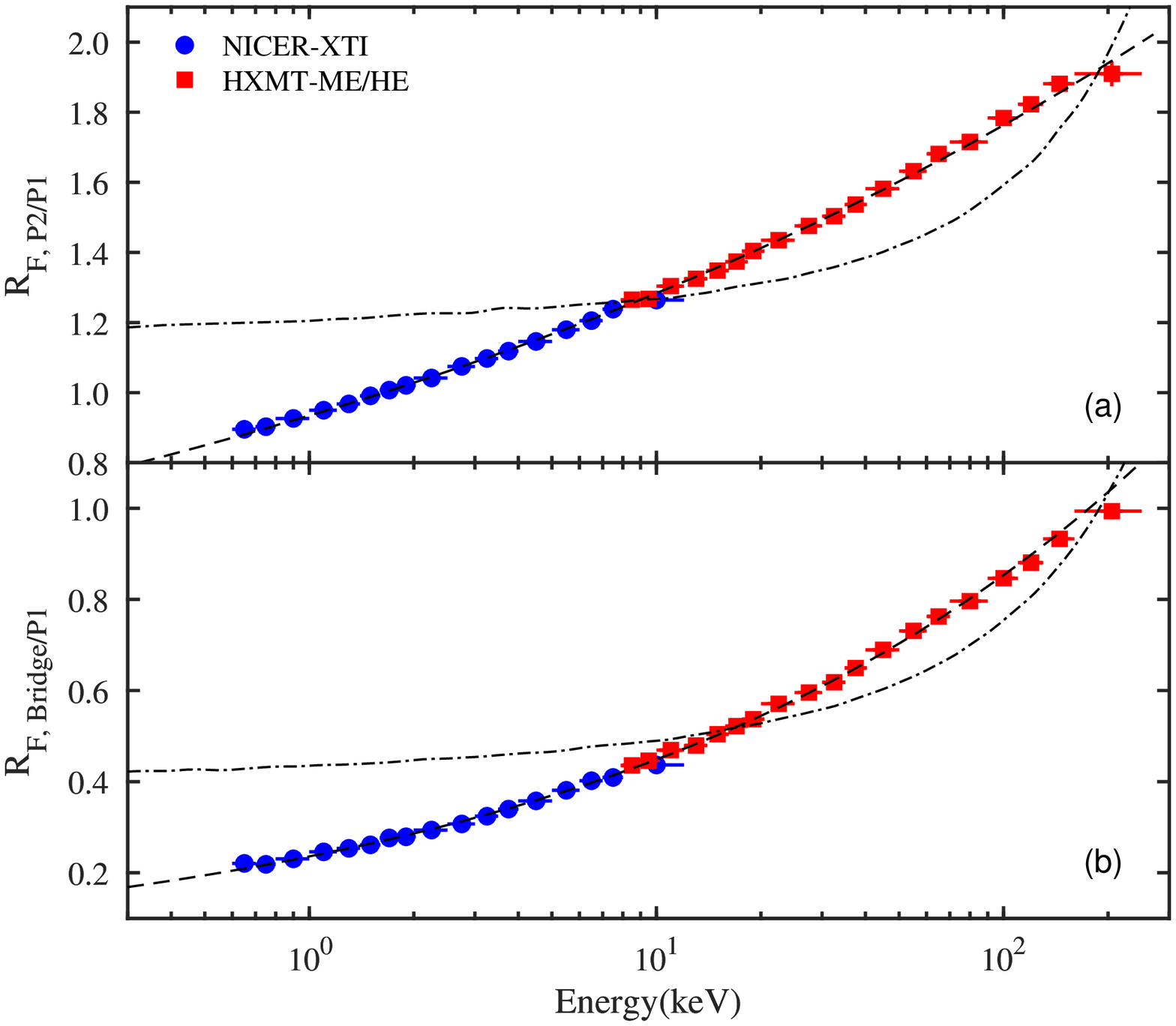}  
\caption{The flux ratios of the Crab pulsar using the phase
intervals defined in \citet{Zhang2002}, which
are (-0.05,0.05) for P1, (0.05,0.27) for Bridge, and (0.27,0.47)
for P2. The dash-dotted lines in panel (a) and (b) are the expected
evolution trends of flux ratios for the model in \citet{Zhang2002}.
The dashed lines represent the evolution trends of flux ratios from
observations.
\label{flux_ratio_zhang}}
\end{center}
\end{figure}

In order to compare our observational results with the theoretic
prediction of the outer gap model in \citet{Zhang2002}, flux ratios
($R_{F, P2/P1}$, $R_{F, Bridge/P1}$) are calculated according to the
phase intervals used in their work, which are (-0.05, 0.05)
for P1, (0.05, 0.27) for Bridge, and (0.27, 0.47) for P2.
Apparently, as shown in Figure \ref{flux_ratio_zhang}, \citet{Zhang2002}'s
model can produce qualitatively the observed trend, but their exact
functional relationship between flux ratio and energy is not accord with
the results in this work.

\subsection{Phase-resolved Spectra}
\begin{figure}
\begin{center}
\includegraphics[angle=0, width=0.65\textwidth]{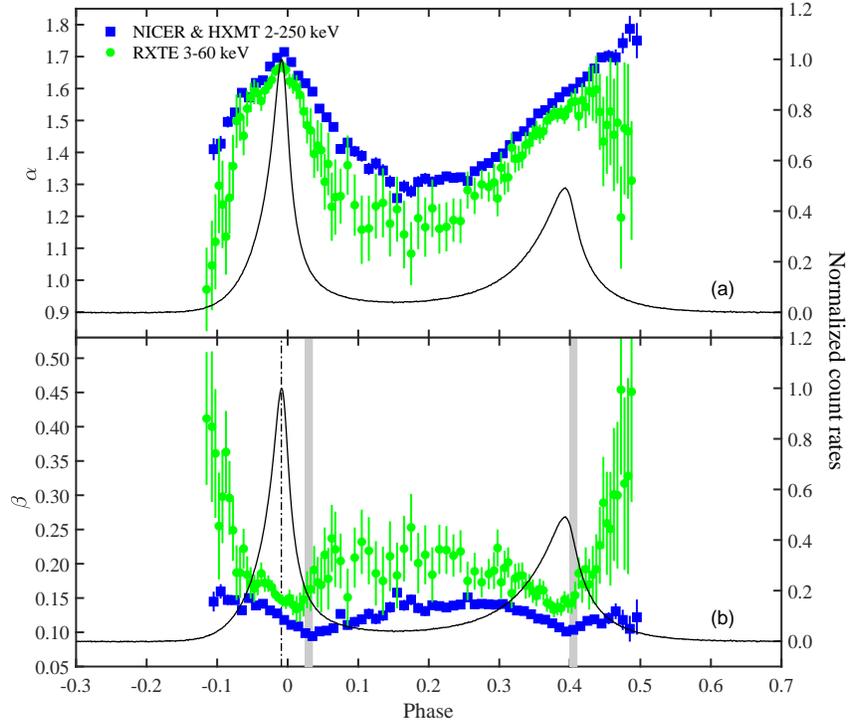}
\caption{Fitting results of the PRS of the Crab pulsar using a
log-parabola model, where these blue squared data points are the joint
fitting results from {\sl NICER} and {\sl Insight-HXMT}'s observations
which cover 2-250 keV, and the green circled data points are the fitting
results for {\sl RXTE}'s observations presented in \citet{Ge2012ApJS}.
The dash-dotted line marks the peak location of the primary pulse P1,
and the two gray bars mark the shape turnover phase intervals, as shown
in Figure \ref{diff_energy_profiles}.
\label{spectra_index}}
\end{center}
\end{figure}

The fitted parameters of the PRS by the log-parabola
model are displayed in Figure \ref{spectra_index}, and their values
and fitting goodness are listed in Table \ref{table:2}.
As can be seen in Figure \ref{spectra_index}(a), the phase resolved
spectral properties of the two pulses are quite different: the spectral
index $\alpha$ around P1 increases first and then falls off, with
the maximum at the peak; and $\alpha$ keeps growing from the leading
wing to the trailing wing of P2. In the bridge region, $\alpha$ has a
down and up trend, and changes smoothly with phase.
The values of the curvature parameter $\beta$ for all the
spectra are positive ( Figure \ref{spectra_index}(b) ), meaning that
the spectral index increases with energy at all phases. $\beta$ presents
a ``W" shape in phase and reaches the local minimum around two turning
points of the pulse profiles.
Both $\alpha$ and $\beta$ in 2-250 keV show phase evolution trends
similar to those in \citet{Ge2012ApJS}, but with higher precision.
The value of $\beta$ in 2-250 keV is smaller than in 3-60 keV, indicating
that the photon index changes more slowly at higher energies.
The phase turning points of $\beta$ in this work (gray phase
regions in Figure \ref{spectra_index}(b)) are not aligned with
those in \citet{Ge2012ApJS}, probably due to the spectral complexity
of this pulsar, because the energy ranges of the two studies are different.

\begin{figure}
\begin{center}
\includegraphics[angle=0,width=0.6\textwidth]{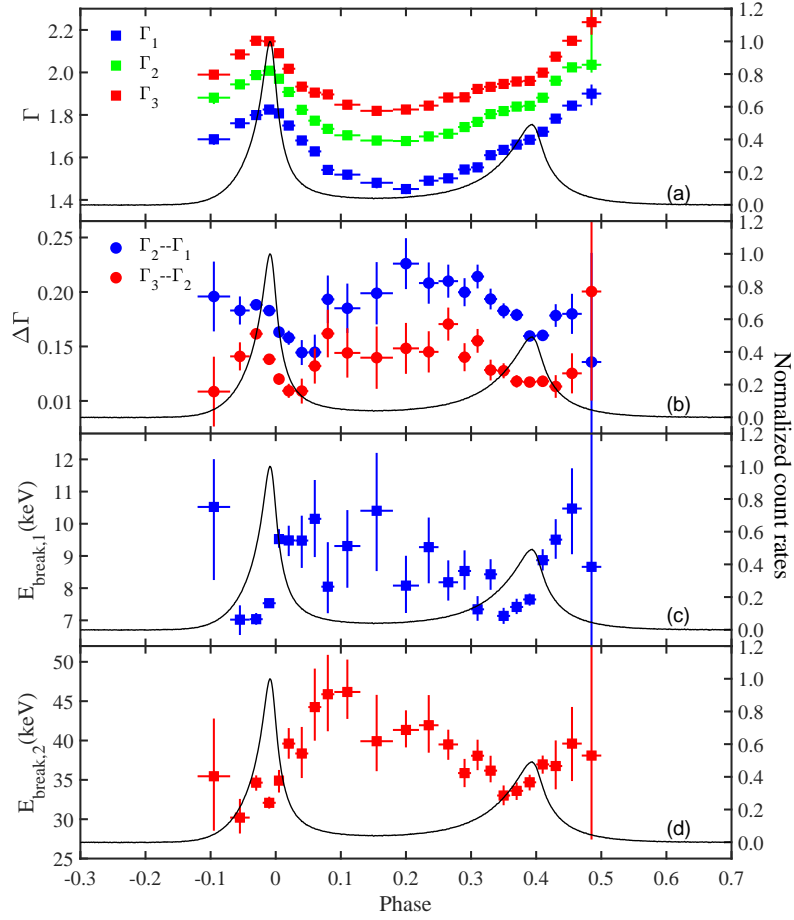}  
\caption{Joint fitting results of the PRS of the Crab pulsar using a three-segment
broken power law model (\textit{bkn2pow}) for X-ray observations in
2--250\,keV. Panel (a),(c),(d): the parameters in the \textit{bkn2pow}
model, and panel (b) displays the differences among three indices
in panel (a).
\label{spectra_index02}}
\end{center}
\end{figure}

As shown in Figure \ref{fig:spec_3models}, in phase interval
0-0.01, the fitting reduced $\chi^2$ are 1.05, 1.23, 1.03 when
using $logpar$, $bknpow$, and $bkn2pow$ models, respectively.
The two-segment broken power law model does not result in
statistically acceptable fit to the broad band PRS,
whereas a three-segment broken power law model can fit the spectra very well.
The joint fitting parameters for PRS by using a three-segment broken
power law are shown in Figure \ref{spectra_index02} and listed
in Table \ref{table:3}. It can be seen that the three spectral
indices ($\Gamma_{1}$, $\Gamma_{2}$, $\Gamma_{3}$) all
change with phase and their evolution trends are similar, and
distributions of the differences between them are consistent with
the evolution of $\beta$ with phase. $E_{break,1}$ changes rather
randomly, but $E_{break,2}$ is positively correlated with $\beta$.
\cite{Madsen2015APJ} found that the 3-78 keV PRS extracted from
the {\sl NuSTAR} observations can be well fitted with a two-segment
power-law function (\textit{bknpow}), with the break energy fixed at
11.7 keV or 13.1 keV. Comparing with our results, we found that these
two break energies are comparable with $E_{break,1}$ and the evolutions
of the photon indices share the similar trend.
According to the fitting results in Table \ref{table:3}, the reduced
$\chi^2$ is higher and closer to 1 in the phase intervals near P1 and
P2 than that in the bridge intervals, it means the \textit{bkn2pow}
model is more suitable for the spectra fitting of P1 and P2, but the
spectra in the other phase intervals with lower statistics could be
over fitted with such a model.
It should be noted that the NuSTAR results \citep{Madsen2015APJ} also
support our results that the turnover or bending at high energies are
around a few tens of keV.

\begin{figure}
\begin{center}
\includegraphics[angle=0,width=1.0\textwidth]{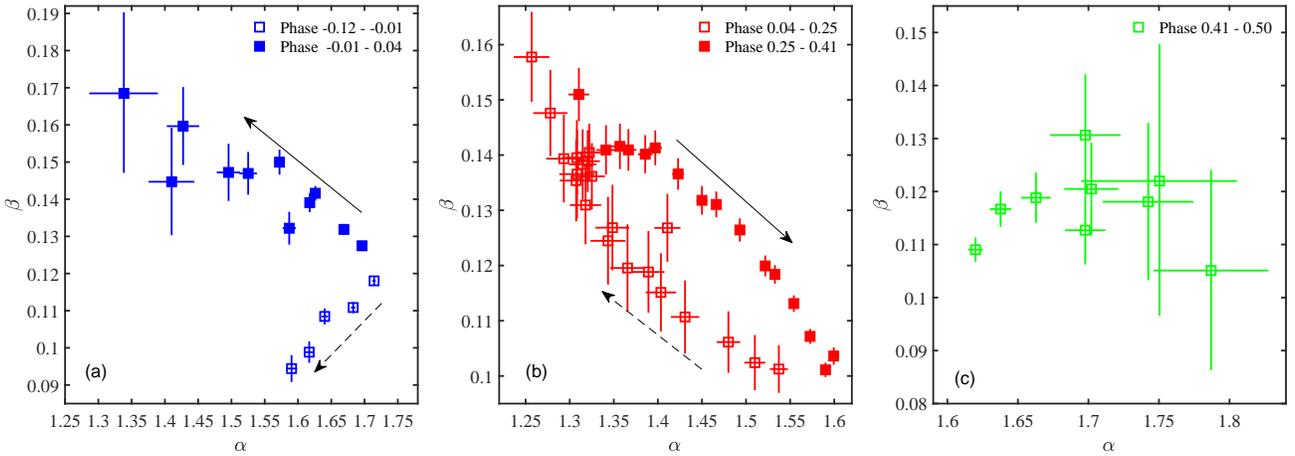}
\caption{The correlations between $\alpha$ and $\beta$ in different
phase intervals. The black dashed and solid arrows indicate the
direction of phase increasing for their surrounding hollow and solid
data points, respectively.
\label{correlation01}}
\end{center}
\end{figure}

\begin{figure}
\begin{center}
\includegraphics[angle=0,width=0.5\textwidth]{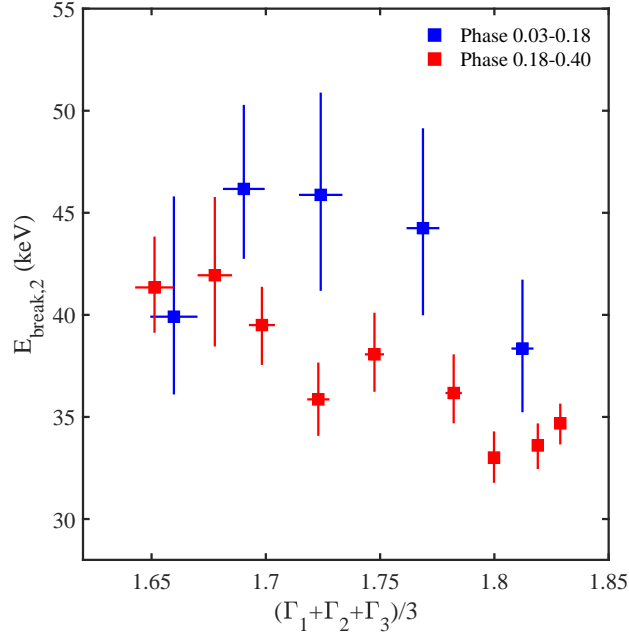}
\caption{The correlations between the average photon indices and $E_{break,2}$
of the spectral model $bkn2pow$ in different phase intervals.
\label{correlation02}}
\end{center}
\end{figure}

Some fitting parameters of \textit{logpar} and \textit{bkn2pow} models
have clear correlations in certain phase intervals. As shown in
Figure \ref{correlation01}(a), the correlation between $\alpha$ and
$\beta$ are opposite on both sides of P1 before reaching the first
phase turning point. Between the two phase turning points, there is a
significant anti-correlation between $\alpha$ and $\beta$, but their
functional relationship seem to be not the same for two groups of data,
in phase 0.04-0.25 and 0.25-0.41 respectively (Figure \ref{correlation01}(b)).
At the right side of P2, the correlation between $\alpha$ and
$\beta$ is unclear, perhaps due to the limitation of statistics
(Figure \ref{correlation01}(c)).
For the \textit{bkn2pow} model, we only find a possible anti-correlation
between the averaged indices (($\Gamma_{1}+\Gamma_{2}+\Gamma_{3}$)/3) and
$E_{break,2}$ in phase 0.03-0.40, and their exact functional relationship
also seems to be not the same in the front and rear phase intervals of
the bridge region, as shown in Figure \ref{correlation02}.

\begin{table}
\footnotesize
\caption{The Phase-resolved Spectral Analysis Results of the
Crab Pulsar for a Log-parabola Model (\textit{logpar}) in 2-250 keV}
\scriptsize{}
\label{table:2}
\medskip
\begin{center}
\begin{tabular}{c c c c c}
\hline \hline
Phase Range & $\alpha$ & $\beta$ & Normalization & Reduced $\chi^2$ \\
\hline
   0 - 0.01	& $1.683\pm0.003$ & $0.111\pm0.001$ & $2.512\pm0.008$ &	1.05 	\\
0.01 - 0.02	& $1.640\pm0.005$ & $0.108\pm0.002$ & $1.414\pm0.007$ &	1.04 	\\
0.02 - 0.03	& $1.617\pm0.007$ & $0.099\pm0.003$ & $0.913\pm0.007$ &	1.01 	\\
0.03 - 0.04	& $1.590\pm0.008$ & $0.094\pm0.004$ & $0.643\pm0.006$ &	0.99 	\\
0.04 - 0.05	& $1.537\pm0.010$ & $0.101\pm0.004$ & $0.470\pm0.005$ &	1.02 	\\
0.05 - 0.06	& $1.510\pm0.012$ & $0.102\pm0.005$ & $0.372\pm0.005$ &	1.03 	\\
0.06 - 0.07	& $1.480\pm0.013$ & $0.106\pm0.006$ & $0.308\pm0.005$ &	1.01 	\\
0.07 - 0.08	& $1.411\pm0.015$ & $0.127\pm0.006$ & $0.250\pm0.005$ &	1.01 	\\
0.08 - 0.09	& $1.431\pm0.016$ & $0.111\pm0.007$ & $0.225\pm0.004$ &	1.04 	\\
0.09 - 0.10	& $1.403\pm0.017$ & $0.115\pm0.007$ & $0.199\pm0.004$ &	1.06 	\\
0.10 - 0.11	& $1.389\pm0.018$ & $0.119\pm0.007$ & $0.184\pm0.004$ &	0.97 	\\
0.11 - 0.12	& $1.348\pm0.019$ & $0.127\pm0.008$ & $0.162\pm0.004$ &	1.05 	\\
0.12 - 0.13	& $1.366\pm0.020$ & $0.120\pm0.008$ & $0.161\pm0.004$ &	0.99 	\\
0.13 - 0.14	& $1.343\pm0.020$ & $0.124\pm0.008$ & $0.154\pm0.004$ &	1.04 	\\
0.14 - 0.15	& $1.308\pm0.020$ & $0.137\pm0.008$ & $0.145\pm0.004$ &	1.00 	\\
0.15 - 0.16	& $1.257\pm0.020$ & $0.158\pm0.008$ & $0.137\pm0.004$ &	1.01 	\\
0.16 - 0.17	& $1.293\pm0.020$ & $0.139\pm0.008$ & $0.142\pm0.004$ &	1.00 	\\
0.17 - 0.18	& $1.278\pm0.019$ & $0.148\pm0.008$ & $0.145\pm0.004$ &	1.01 	\\
0.18 - 0.19	& $1.307\pm0.018$ & $0.135\pm0.007$ & $0.156\pm0.004$ &	0.98 	\\
0.19 - 0.20	& $1.318\pm0.018$ & $0.131\pm0.007$ & $0.164\pm0.004$ &	1.00 	\\
0.20 - 0.21	& $1.308\pm0.017$ & $0.140\pm0.007$ & $0.175\pm0.004$ &	0.97 	\\
0.21 - 0.22	& $1.315\pm0.016$ & $0.138\pm0.006$ & $0.191\pm0.004$ &	1.02 	\\
0.22 - 0.23	& $1.325\pm0.015$ & $0.136\pm0.006$ & $0.206\pm0.004$ &	1.00 	\\
0.23 - 0.24	& $1.320\pm0.014$ & $0.139\pm0.006$ & $0.222\pm0.004$ &	0.98 	\\
0.24 - 0.25	& $1.322\pm0.013$ & $0.140\pm0.005$ & $0.242\pm0.004$ &	1.02 	\\
0.25 - 0.26	& $1.310\pm0.012$ & $0.151\pm0.005$ & $0.266\pm0.004$ &	1.00 	\\
0.26 - 0.27	& $1.341\pm0.011$ & $0.141\pm0.004$ & $0.303\pm0.004$ &	1.04 	\\
0.27 - 0.28	& $1.357\pm0.010$ & $0.142\pm0.004$ & $0.348\pm0.004$ &	1.05 	\\
0.28 - 0.29	& $1.367\pm0.009$ & $0.141\pm0.004$ & $0.393\pm0.004$ &	1.06 	\\
0.29 - 0.30	& $1.386\pm0.008$ & $0.140\pm0.003$ & $0.456\pm0.005$ &	1.00 	\\
0.30 - 0.31	& $1.397\pm0.008$ & $0.141\pm0.003$ & $0.523\pm0.005$ &	1.00 	\\
0.31 - 0.32	& $1.423\pm0.007$ & $0.137\pm0.003$ & $0.615\pm0.005$ &	0.95 	\\
0.32 - 0.33	& $1.450\pm0.006$ & $0.132\pm0.003$ & $0.726\pm0.005$ &	0.98 	\\
0.33 - 0.34	& $1.466\pm0.006$ & $0.131\pm0.002$ & $0.851\pm0.005$ &	1.05 	\\
0.34 - 0.35	& $1.493\pm0.005$ & $0.126\pm0.002$ & $1.017\pm0.006$ &	1.02 	\\
0.35 - 0.36	& $1.522\pm0.004$ & $0.120\pm0.002$ & $1.218\pm0.006$ &	1.02 	\\
0.36 - 0.37	& $1.533\pm0.004$ & $0.118\pm0.002$ & $1.426\pm0.006$ &	1.05 	\\
0.37 - 0.38	& $1.554\pm0.003$ & $0.113\pm0.001$ & $1.699\pm0.006$ &	1.12 	\\
0.38 - 0.39	& $1.573\pm0.003$ & $0.107\pm0.001$ & $1.966\pm0.007$ &	1.01 	\\
0.39 - 0.40	& $1.590\pm0.003$ & $0.101\pm0.001$ & $2.072\pm0.007$ &	1.07 	\\
0.40 - 0.41	& $1.599\pm0.004$ & $0.104\pm0.002$ & $1.739\pm0.007$ &	1.07 	\\
0.41 - 0.42	& $1.620\pm0.005$ & $0.109\pm0.002$ & $1.248\pm0.007$ &	1.01 	\\
0.42 - 0.43	& $1.638\pm0.008$ & $0.117\pm0.003$ & $0.901\pm0.007$ &	0.97 	\\
0.43 - 0.44	& $1.663\pm0.010$ & $0.119\pm0.005$ & $0.683\pm0.007$ &	1.02 	\\
0.44 - 0.45	& $1.698\pm0.014$ & $0.113\pm0.006$ & $0.530\pm0.008$ &	0.94 	\\
0.45 - 0.46	& $1.702\pm0.019$ & $0.120\pm0.009$ & $0.407\pm0.008$ &	0.96 	\\
0.46 - 0.47	& $1.698\pm0.025$ & $0.131\pm0.011$ & $0.320\pm0.008$ &	1.09 	\\
0.47 - 0.48	& $1.742\pm0.032$ & $0.118\pm0.015$ & $0.260\pm0.008$ &	1.08 	\\
0.48 - 0.49	& $1.787\pm0.041$ & $0.105\pm0.019$ & $0.217\pm0.009$ &	1.02 	\\
0.49 - 0.50	& $1.750\pm0.055$ & $0.122\pm0.026$ & $0.159\pm0.009$ &	1.00 	\\
0.88 - 0.89	& $1.338\pm0.052$ & $0.168\pm0.022$ & $0.070\pm0.004$ &	1.03 	\\
0.89 - 0.90	& $1.410\pm0.034$ & $0.145\pm0.014$ & $0.114\pm0.005$ &	1.05 	\\
0.90 - 0.91	& $1.427\pm0.024$ & $0.160\pm0.010$ & $0.182\pm0.005$ &	1.05 	\\
0.91 - 0.92	& $1.496\pm0.018$ & $0.147\pm0.008$ & $0.287\pm0.006$ &	1.01 	\\
0.92 - 0.93	& $1.525\pm0.013$ & $0.147\pm0.006$ & $0.423\pm0.006$ &	1.03 	\\
0.93 - 0.94	& $1.587\pm0.010$ & $0.132\pm0.004$ & $0.628\pm0.007$ &	1.11 	\\
0.94 - 0.95	& $1.572\pm0.008$ & $0.150\pm0.003$ & $0.870\pm0.007$ &	1.05 	\\
0.95 - 0.96	& $1.618\pm0.006$ & $0.139\pm0.003$ & $1.271\pm0.007$ &	1.01 	\\
0.96 - 0.97	& $1.626\pm0.004$ & $0.142\pm0.002$ & $1.784\pm0.008$ &	1.13 	\\
0.97 - 0.98	& $1.669\pm0.003$ & $0.132\pm0.001$ & $2.658\pm0.009$ &	1.08 	\\
0.98 - 0.99	& $1.696\pm0.002$ & $0.127\pm0.001$ & $3.898\pm0.010$ &	1.12 	\\
0.99 - 1.00	& $1.714\pm0.002$ & $0.118\pm0.001$ & $4.226\pm0.010$ &	1.20 	\\
\hline \hline
\end{tabular}
\end{center}
\end{table}

\begin{table}
\footnotesize
\caption{The Phase-resolved Spectral Analysis Results of the
Crab Pulsar for a Three-segment Broken Function Model (\textit{bkn2pow})
in 2-250 keV}
\scriptsize{}
\label{table:3}
\medskip
\begin{center}
\begin{tabular}{c c c c c c c c c}
\hline \hline
Phase Range & $\Gamma_{1}$ & $\Gamma_{2}$ & $\Gamma_{3}$ & $E_{break,1}$
& $E_{break,2}$ & Normalization & Reduced $\chi^2$ \\
\hline
0 - 0.01 & $1.807_{-0.003}^{+0.003}$ & $1.970_{-0.004}^{+0.004}$ & $2.090_{-0.005}^{+0.005}$ & $9.523_{-0.269}^{+0.308}$ & $34.907_{-1.516}^{+1.353}$ & $2.663_{-0.009}^{+0.009}$ & 1.03 	\\
0.01 - 0.03 & $1.750_{-0.005}^{+0.004}$ & $1.908_{-0.005}^{+0.004}$ & $2.017_{-0.006}^{+0.005}$ & $9.482_{-0.486}^{+0.459}$ & $39.610_{-1.922}^{+1.970}$ & $1.236_{-0.007}^{+0.007}$ & 0.75 	\\
0.03 - 0.05 & $1.680_{-0.008}^{+0.008}$ & $1.824_{-0.008}^{+0.008}$ & $1.933_{-0.008}^{+0.010}$ & $9.478_{-0.850}^{+0.773}$ & $38.351_{-3.120}^{+3.376}$ & $0.592_{-0.006}^{+0.007}$ & 0.73 	\\
0.05 - 0.07 & $1.628_{-0.013}^{+0.012}$ & $1.773_{-0.011}^{+0.010}$ & $1.905_{-0.014}^{+0.015}$ & $10.154_{-1.188}^{+1.206}$ & $44.248_{-4.268}^{+4.892}$ & $0.369_{-0.006}^{+0.006}$ & 0.76 	\\
0.07 - 0.09 & $1.541_{-0.017}^{+0.021}$ & $1.734_{-0.010}^{+0.010}$ & $1.896_{-0.017}^{+0.019}$ & $8.046_{-0.825}^{+1.382}$ & $45.878_{-4.701}^{+5.009}$ & $0.253_{-0.005}^{+0.007}$ & 0.69 	\\
0.09 - 0.13 & $1.519_{-0.023}^{+0.019}$ & $1.704_{-0.008}^{+0.009}$ & $1.848_{-0.014}^{+0.017}$ & $9.308_{-1.296}^{+1.117}$ & $46.170_{-3.425}^{+4.116}$ & $0.192_{-0.005}^{+0.005}$ & 0.60 	\\
0.13 - 0.18 & $1.481_{-0.026}^{+0.022}$ & $1.679_{-0.013}^{+0.018}$ & $1.819_{-0.010}^{+0.014}$ & $10.404_{-1.874}^{+1.798}$ & $39.911_{-3.808}^{+5.892}$ & $0.163_{-0.006}^{+0.005}$ & 0.58 	\\
0.18 - 0.22 & $1.451_{-0.024}^{+0.020}$ & $1.677_{-0.008}^{+0.007}$ & $1.825_{-0.010}^{+0.010}$ & $8.079_{-0.851}^{+0.926}$ & $41.344_{-2.218}^{+2.491}$ & $0.185_{-0.006}^{+0.005}$ & 0.55 	\\
0.22 - 0.25 & $1.491_{-0.017}^{+0.016}$ & $1.699_{-0.010}^{+0.009}$ & $1.844_{-0.011}^{+0.014}$ & $9.274_{-1.123}^{+0.920}$ & $41.941_{-3.489}^{+3.832}$ & $0.249_{-0.005}^{+0.006}$ & 0.59 	\\
0.25 - 0.28 & $1.501_{-0.015}^{+0.013}$ & $1.711_{-0.007}^{+0.007}$ & $1.882_{-0.008}^{+0.008}$ & $8.182_{-0.569}^{+0.681}$ & $39.501_{-1.953}^{+1.876}$ & $0.337_{-0.006}^{+0.006}$ & 0.64 	\\
0.28 - 0.30 & $1.543_{-0.011}^{+0.011}$ & $1.743_{-0.006}^{+0.007}$ & $1.883_{-0.007}^{+0.007}$ & $8.533_{-0.584}^{+0.641}$ & $35.861_{-1.785}^{+1.806}$ & $0.471_{-0.007}^{+0.006}$ & 0.75 	\\
0.30 - 0.32 & $1.553_{-0.008}^{+0.011}$ & $1.767_{-0.005}^{+0.006}$ & $1.922_{-0.006}^{+0.007}$ & $7.342_{-0.356}^{+0.410}$ & $38.067_{-1.835}^{+2.039}$ & $0.615_{-0.006}^{+0.008}$ & 0.70 	\\
0.32 - 0.34 & $1.610_{-0.008}^{+0.008}$ & $1.804_{-0.005}^{+0.005}$ & $1.932_{-0.005}^{+0.005}$ & $8.429_{-0.481}^{+0.469}$ & $36.172_{-1.485}^{+1.893}$ & $0.865_{-0.008}^{+0.008}$ & 0.77 	\\
0.34 - 0.36 & $1.635_{-0.006}^{+0.005}$ & $1.818_{-0.004}^{+0.004}$ & $1.946_{-0.004}^{+0.004}$ & $7.138_{-0.249}^{+0.251}$ & $33.010_{-1.236}^{+1.281}$ & $1.200_{-0.009}^{+0.008}$ & 0.74 	\\
0.36 - 0.38 & $1.660_{-0.004}^{+0.005}$ & $1.839_{-0.004}^{+0.003}$ & $1.957_{-0.003}^{+0.003}$ & $7.416_{-0.221}^{+0.255}$ & $33.604_{-1.158}^{+1.079}$ & $1.664_{-0.008}^{+0.009}$ & 0.79 	\\
0.38 - 0.40 & $1.683_{-0.003}^{+0.003}$ & $1.843_{-0.003}^{+0.003}$ & $1.960_{-0.003}^{+0.003}$ & $7.648_{-0.193}^{+0.184}$ & $34.687_{-1.034}^{+0.966}$ & $2.134_{-0.008}^{+0.008}$ & 0.74 	\\
0.40 - 0.42 & $1.721_{-0.005}^{+0.004}$ & $1.881_{-0.003}^{+0.003}$ & $1.999_{-0.004}^{+0.004}$ & $8.867_{-0.320}^{+0.343}$ & $36.959_{-1.174}^{+1.129}$ & $1.585_{-0.009}^{+0.008}$ & 0.72 	\\
0.42 - 0.44 & $1.783_{-0.007}^{+0.007}$ & $1.961_{-0.007}^{+0.008}$ & $2.074_{-0.009}^{+0.010}$ & $9.506_{-0.595}^{+0.636}$ & $36.757_{-2.960}^{+3.254}$ & $0.853_{-0.008}^{+0.007}$ & 0.69 	\\
0.44 - 0.47 & $1.844_{-0.014}^{+0.012}$ & $2.024_{-0.013}^{+0.014}$ & $2.149_{-0.020}^{+0.020}$ & $10.473_{-1.419}^{+1.250}$ & $39.611_{-4.762}^{+4.655}$ & $0.456_{-0.007}^{+0.007}$ & 0.62 	\\
0.47 - 0.50 & $1.900_{-0.055}^{+0.043}$ & $2.036_{-0.037}^{+0.138}$ & $2.237_{-0.059}^{+0.639}$ & $8.662_{-2.993}^{+15.236}$ & $38.089_{-10.670}^{+108.327}$ & $0.231_{-0.013}^{+0.013}$ & 0.66 	\\
0.88 - 0.93 & $1.685_{-0.026}^{+0.016}$ & $1.881_{-0.030}^{+0.019}$ & $1.990_{-0.014}^{+0.016}$ & $10.522_{-2.271}^{+1.485}$ & $35.456_{-6.917}^{+7.339}$ & $0.249_{-0.007}^{+0.006}$ & 0.59 	\\
0.93 - 0.96 & $1.761_{-0.010}^{+0.010}$ & $1.944_{-0.009}^{+0.008}$ & $2.085_{-0.006}^{+0.006}$ & $7.022_{-0.478}^{+0.444}$ & $30.206_{-2.023}^{+2.375}$ & $1.023_{-0.011}^{+0.011}$ & 0.66 	\\
0.96 - 0.98 & $1.799_{-0.004}^{+0.003}$ & $1.988_{-0.003}^{+0.003}$ & $2.149_{-0.004}^{+0.004}$ & $7.036_{-0.172}^{+0.192}$ & $34.647_{-1.024}^{+0.926}$ & $2.424_{-0.010}^{+0.010}$ & 0.81 	\\
0.98 - 1.00 & $1.825_{-0.002}^{+0.003}$ & $2.008_{-0.003}^{+0.002}$ & $2.146_{-0.003}^{+0.003}$ & $7.532_{-0.130}^{+0.135}$ & $32.085_{-0.738}^{+0.774}$ & $4.354_{-0.012}^{+0.013}$ & 0.82 	\\
\hline \hline
\end{tabular}
\end{center}
\end{table}

\section{Discussion}
\label{sect:discussion}
\subsection{Comparisons with the Previous Works}
The evolution of the flux ratio of the two pulses, one of
the most frequently used parameters to describe the pulse profile
evolution in X-rays, is consistent with the previous work
in \citet{Kuiper2001AA} and \citet{Tuo2019RAA}, but the current
results are much more precise and cover a wider energy range
of 0.4-250 keV.
A multicomponent model in \citet{Massaro2006AA} is able
to reproduce the pulse profiles in different energy bands and the
evolution of the flux ratio between P2 and P1 \citep{Kuiper2001AA}.
The flux ratio of two pulses in 0.4-250 keV in this work coincides
with that in \citet{Kuiper2001AA} when using the same phase definitions
for P1 and P2, so our results can be also explained by the
multicomponent model in \cite{Massaro2006AA}.

\citet{Zhang2002} once fit the flux ratios of P2 to P1 and Bridge to P1 by
using an outer gap model. However, as shown in Figure \ref{flux_ratio_zhang},
there are obvious discrepancies between the predictions of their model
and our results. The three-dimensional model in \citet{Zhang2002} is based
on a rotating vacuum dipole, in which some processes are simplified or
unreliable. For example, only photons emitting tangent to the local
magnetic field lines are considered, but particles with large initial pitch
angles may also emit in other directions. The magnetic inclination
and viewing angles used in their work are from old theoretical works,
which have not been well determined by observations and are therefore probably
inaccurate. More over, the hard X-ray spectral indices they adopted are
from old observations and with large uncertainties. These factors may lead
to the discrepancies between model predictions and observational results.
Therefore, more details need to be added and/or refined into the
outer gap model of \cite{Zhang2002} to explain the new results in this work.

There are several studies on the spectrum of the Crab pulsar
as a function of phase \citep{Pravdo1997, Massaro2000AA, Massaro2006AA,
Mineo2006AA, Ge2012ApJS, Tuo2019RAA, Li2019, Paul2020}.
When fitting the spectrum by using a power law model, the overall
behaviors of photon indices are very similar, and indices increase clearly
with energy over all the phase intervals, i.e., the higher
the energy, the softer the spectra are. For the energy curvature
trend of spectra which is described by $\beta$ of the log-parabola
model fitting, this work gives a more detailed phase evolution trend
and obtains links between the spectrum curvature and the evolution
of pulse profiles with energy, as shown in Figure \ref{diff_energy_profiles}(b)
and Figure \ref{spectra_index}. The uncertainties of $\alpha$ and
$\beta$ are about one fifth of those in \citet{Ge2012ApJS}.

\subsection{Constraints on Pulsar Radiation Models}
We can use the energy dependence of the pulse profiles and the PRS
to test the pulsar high energy emission models.
One of the most popular models in the past is the two-gap outer
gap model, proposed by Cheng, Ho, and Ruderman \citep{Cheng1986a, Cheng1986b}.
In this model, the gap is ``stratified" in energy, and lower energy
radiation emanates from a region farther from the neutron star surface
than higher energy radiation. The delay between the radio and
X-ray pulses decreases significantly with increasing X-ray energy,
which indicates that the softer X-ray photons originate from higher
altitudes \citep{Molkov2010}. Qualitatively, the energy evolution trends
of the pulse shape and spectral index of the Crab pulsar obtained in this
paper all conform to the ``stratified" structure of this outer gap model.
The gap closer to the neutron star surface may cover a larger angular
size (relative to the neutron star center) as indicated by the increasing
widths of the two pulses with energy.
Because the magnetic field in deeper magnetosphere is stronger
and P2 becomes more prominent at higher energy, the particle generation (and
thus synchrotron emission) in the region close to the neutron star
surface and corresponds to P2 maybe more intense \citep{Cheng2000}.
According to the energy evolution trend of the whole pulse profiles,
the pulse intensity of P1 changes more distinctly than P2, and
corresponding radiation area of P1 may be longer and narrower than P2.

In recent years, global PIC simulations were used to model the
pulsar magnetosphere (such as \cite{Philippov2014, Cerutti2016MNRAS, Philippov2018ApJ}).
In these simulations, the high energy emissions are mainly produced
in current sheets just beyond the light cylinder.
\citet{Cerutti2016MNRAS} modeled the pulse profiles and phase-averaged
spectra produced by synchrotron radiation process at different
viewing and magnetic inclination angles. When the viewing angle is
$\sim$ 90-100$^\circ$ and the magnetic inclination is $\sim$ 30-45$^\circ$,
the model predicted spectra and pulse profiles of the Crab pulsar
are well consistent with the observational results in energy band from
soft X-ray to $\gamma$-ray as presented in \cite{Kuiper2001AA}.
The PIC simulations and modeling show that electrons and positrons
are generated mainly within the magnetosphere. Near the pulsar surface,
the electric potential is highest in the equatorial plane and decreases
toward the poles for a quasi-aligned rotator \citep{GJ1969ApJ}. The freshly
generated positrons in the magnetosphere are accelerated outward to high
energy in the reconnection regions outside the light cylinder, where electrons
coming from farther in the equatorial plane are accelerated inward.
Because the supply of outgoing positrons is much higher than the ingoing
electrons, high energy radiation is mainly generated by positrons for a
quasi-aligned rotator \citep{Cerutti2016MNRAS}, and on the contrary, for an
anti-aligned rotator, it is expected that the high energy radiation is
generated mainly by electrons. The highest energy that a positron can acquire
is determined by the accelerating electric field in the reconnection region,
multiplied by the acceleration distance that is available in the undulating
current sheet. As a consequence,  the spectrum will be harder for a lower
magnetic inclination angle. On the other hand, charged particles in the
current sheet can screen the accelerating electric field \citep{Philippov2018ApJ}.
These combined factors imply that the high intensity of P1 in low energy
band is generated by regions where positrons are more numerous and
therefore electric yielding is more significant. In higher energy band,
the intensity of P2 becomes higher because P2 is
likely produced by regions where electric yielding is weak and positrons
can be accelerated to higher energy. The spectra of the bridge component
are harder than those of both P1 and P2, while its intensity is lower
than P1 and P2. This behavior can also be explained if the bridge component
is generated by regions where positron density is lower and the fraction
of high energy positron is higher. Future comparison between simulations
of the current sheet in different energy bands with the results in our
paper could provide more detailed information about the particle acceleration
and radiation in the reconnection regions.

\citet{Weisskopf2011ApJ} reported that the soft X-ray spectral
index has an abrupt change in phase 0.83-0.95, where transition
structures of the optical and X-ray polarization properties were also
detected \citep{Slowikowska2009, Vadawale2018Nat}.
We find that around that phase the break energies of the
\textit{bkn2pow} are the lowest. It is yet unknown whether these
properties are physically associated. Besides this, just behind
the two pulses the polarization degree has two local minimum points,
where the values of $\beta$ are also the lowest.
We speculate that in the two corresponding regions the particles
evolve the least due to the lowest magnetic field, which could be
less regulated and thus the radiations are less polarized.

\section{Summary}
In this work, we adopt the observations from {\sl NICER} and {\sl Insight-HXMT},
and study the pulse profile and spectrum of the Crab pulsar systematically.
The main results and conclusions are: (1) in 0.4-250 keV, the intensity/flux
ratio of two pulses show an increasing trend with energy, so do the widths of
the two pulses; the separation of the two pulses decreases with energy;
the energy evolution trends of the two sides of P1 and P2 are asymmetric;
(2) in 2-250 keV, where the spectral shape is not significantly
affected by the chemical composition of the absorption interstellar
medium, the spectral index increases with energy over all the phase
intervals; the spectral index changes more slowly at higher energies;
the change of spectral index reaches its local minimum around
phase 0.035 and 0.4 that are later than the peak phase of P1 and P2;
there are obvious energy breaks on the spectra near the two peaks of the
Crab pulsar, and the break energies are generally higher in the region
bridging the two pulses;
(3) the energy evolution trends of pulse profile and spectrum conform to
results in previous studies qualitatively, but more accurate to allow
for a detailed analysis;
(4) the anti-correlations among spectral fitting parameters ($\alpha$
and $\beta$) indicate that the highest energy particles are produced in
regions where the radiation energy loss is also the strongest.
We simply compared our results with the predictions of the outer gap
model and the PIC simulations of pulsar magnetosphere. It turns out that
the theoretical models need to be further refined to explain the details
of the observational results.

\begin{acknowledgments}
We thank the referee for his/her very helpful comments.
We appreciate Prof. Zhang Li, Jiang Zejun, Tong Hao, Dr. Bu Qingcui
and Xiao Guangcheng for their useful suggestions and discussions.
We thank the High Energy Astrophysics Science Archive Research Center
for maintaining its online archive service that provided the data
used in this research. This work also makes use of the data from the
{\sl Insight-HXMT} mission, a project funded by China National Space
Administration (CNSA) and the Chinese Academy of Sciences (CAS).
This work is supported by the National Key R\&D Program of
China (No. 2021YFA0718500), the National Natural Science Foundation of
China (No.11903001, U1938109, U1838201, U1838202, and 42004004).
We also thank the support from the Strategic Priority Program on
Space Science, the Chinese Academy of Sciences, Grant No.
XDA15020503, and the Key Research Foundation of Education
Ministry of Anhui Province (KJ2019A0787), the Doctor Foundation of
Anhui Jianzhu University 2019 (2019QDZ14).
\end{acknowledgments}

\bibliographystyle{aasjournal}


\end{document}